\documentclass[preprint,fleqn,showpacs]{revtex4}

\usepackage{amsmath}
\usepackage[]{graphicx}

\begin{document}
\title[INFLUENCE OF DIFFUSION OF ATOMS]
{INFLUENCE OF DIFFUSION OF ATOMS ON THE DARK RESONANCE LINESHAPE IN SPATIALLY BOUNDED LASER FIELDS}%
\author{V.I. ROMANENKO}
\affiliation{Institute of Physics, Nat. Acad. of Sci. of Ukraine}
\address{46, Nauky Ave., Kyiv 03680, Ukraine}
\email{vr@iop.kiev.ua}
\author{A.V. ROMANENKO}%
\affiliation{Taras Shevchenko National University of Kyiv}%
\address{2, Academician Glushkov Ave., Kyiv 03022, Ukraine}%
\author{L.P. YATSENKO}
\affiliation{Institute of Physics, Nat. Acad. of Sci. of Ukraine}%
\address{46, Nauky Ave., Kyiv 03680, Ukraine}%
\email{vr@iop.kiev.ua}


\def\pr{^{\,\prime}}
\def\ds{\displaystyle}

\def\erfc{\,\mathrm{erfc}}
\def\erf{\,\mathrm{erf}}
\def\ei{\,\mathrm{Ei}}

\def\const{\,\mathrm{const}}
\def\re{{\,\mathrm{Re}\,}}
\def\im{{\,\mathrm{Im}\,}}

\def\A{\mathcal{A}}
\def\C{\mathrm{C}}
\def\SI{\mathcal{S}}
\def\F{\mathsf{F}}
\def\I{\mathcal{I}}

\def\sh{\,\mathrm{sh}}
\def\ch{\,\mathrm{ch}}
\def\th{\,\mathrm{th}}
\def\laplace{\Delta}

\def\eff{\mathrm{eff}}

\begin{abstract}
We propose a diffusion model for the recently discovered
diffusion-induced Ramsey narrowing arising when atoms diffuse in a
buffer-gas cell in the laser radiation field. The diffusion
equation for the coherence of metastable states coupled with an
excited state by laser radiation of different frequencies in a
three-level scheme of the atom-field interaction is obtained in
the strong-collision approximation. The dependence of the shape of an absorption
line near the transmission maximum of one of the frequencies
on the two-photon resonance detuning for
various geometries of the cell is investigated.
\end{abstract}


\maketitle

\section{Introduction}

\label{introduction}

In a three-level system subjected to two laser fields coupling two
metastable states (or a metastable state and a stable one) with an
excited one, a dark or light-nonabsorbing state, namely a
coherent superposition of two metastable states, can be formed. This phenomenon
is called a coherent population trapping (CPT). The
condition of the formation of a CPT state is the two-photon resonance under
interaction of an atom with light, where the difference between
the frequencies of two laser fields is equal to the frequency of
a transition between metastable states. If this condition is
realized, one observes an abrupt decrease of the fluorescence
intensity of the atom in the laser radiation
field~\cite{Alz76-5,Ari76-333,Gra78-218}. Due to the CPT
phenomenon, it is possible to create a window in the absorption
spectrum. As a result, light can propagate almost without losses
through a medium that absorbs light under usual conditions, which
represents the well-known phenomenon of electromagnetically
induced transparency (EIT)~\cite{Har97}. In addition, dark
resonances are used for the light slowing down~\cite{Hau99} and in
the construction of compact laser frequency standards~\cite{Kna01} and
underlie the effective method of population transfer between
different states of atoms or molecules~-- stimulated Raman
adiabatic passage (STIRAP)~\cite{Ber98}. The resonance width
depends on both the coherence decrease rate for the lower states and on
other factors, particularly on the pattern of atomic motion in a
buffer-gas cell. The latter aspect will be the focus of our
attention in this work.

In the case where atoms move through a laser beam of finite width,
the role of the coherence time is played by the residence time of
an atom in the field. If the cell contains a buffer gas in
addition to active atoms, their residence time in the laser beam
increases. As a result, narrow resonances with a width of the
order of tens of hertzs are registered in buffer-gas
cells~\cite{Erh01}. The authors of works
\cite{Erh01,quivers,arimondo} emphasize the role of the buffer gas
in the experiments on coherent population trapping in a
three-level system, though it is considered that, after atoms have
left the region of interaction with radiation, they do not return
there anymore.

A more detailed description of the process of atom-field
interaction must take into account that, having left the region of
interaction with the field, the atom can return there
again~\cite{zibrov1,zibrov2}, so that atoms can interact with
radiation several times before the loss of coherence. Thus,
diffusion of atoms in a buffer gas essentially affects their
response to the resonance excitation by the laser field. If the
coherence relaxation time of the metastable states considerably
exceeds the time till the repeated atom-field interaction, the
atom can get back without loss of coherence after having spent
some time in a dark region (beyond the beam). As a result, it is
worth expecting the narrowing of the resonance line. This
phenomenon was called a diffusion-induced Ramsey
narrowing~\cite{xiao} (by analogy with the Ramsey method of
separated oscillating fields~\cite{ramsey,Ram90}).

\begin{figure}
\includegraphics[width=5cm]{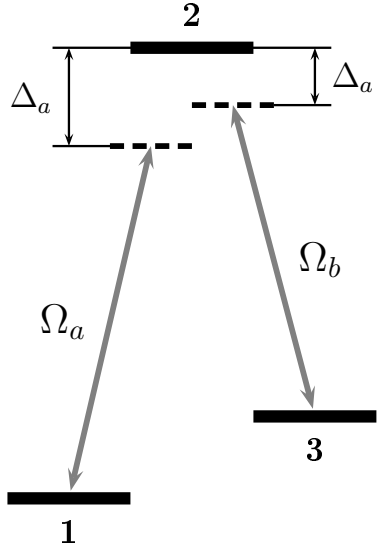}
\caption{Three-level system: $\Omega_{a}$ and $\Omega_{b}$ are the
Rabi frequencies  }
\end{figure}

The diffusion-induced Ramsey narrowing of the transmission spectrum
in the case of EIT observations was investigated in~\cite{xiao}
adducing the results of experiments with rubidium vapor and neon serving
as a buffer gas, as well as the results of theoretical calculations. According to
\cite{xiao}, the necessity of taking
the diffusion process into account depends on the laser-beam diameter~-- the
contribution made into the signal by atoms that had spent some time
beyond the beam and got back becomes determinative with decrease in the
beam diameter. The same authors subsequently published a detailed
description of the developed theory~\cite{xiao2008}. Experimental
results and theoretical calculations unambiguously confirm the
physical interpretation of the phenomenon of diffusion-induced
Ramsey narrowing of the transmission spectrum observed in
\cite{xiao}. In particular, with increase in the laser-beam diameter or
the buffer gas pressure, the shape of the spectral line changes
from the non-Lorentz to Lorentz one in accordance with a decrease
of the contribution made by atoms that have come back from the
region beyond the laser field. In the cited works, the motion of
atoms is described in the form of the Ramsey sequences: in each of
them, an atom spends the time $t^{\rm in}_{1}$ in the radiation
field, moves in the dark region during the time $t^{\rm out}_{2}$,
returns back to the radiation region for the time $t^{\rm in}_{3}$,
and so on. In order to find the transmission spectrum, the density
matrix describing the Ramsey sequence was integrated with the
probability distribution, i.e. one performed averaging over all
possible trajectories. The probability distribution was obtained
in~\cite{xiao2008} from the diffusion equation.

We propose an alternative approach to the description of the
diffusion-induced Ramsey narrowing. It is based on the diffusion
equation obtained from the initial motion equations for the
density matrix on the basis of the strong-collision
approximation~\cite{quivers,sobelman1}. The transmission spectrum
will be directly obtained in the equilibrium approach instead of
averaging over trajectories. As compared to works~\cite{xiao}
and~\cite{xiao2008}, we consider the relaxation at cell
walls (supposing that the coherence is broken due to a collision of an
atom with the wall) and take a realistic (Gaussian) intensity
distribution in the radial direction. In addition, we also
take the finiteness of the gas cell in the direction
parallel to the laser beam into account. In order to understand how the
dimension of the problem influences the result, we also
consider the one-dimensional case where the cell is infinite in
the direction of the laser beam, while the beam itself is infinite
along one of the transverse coordinates.

\section{Basic Equations}
\label{basic-equations}

Let us consider a gas of three-level atoms with the excited state
$\left| 2\right\rangle$ and the metastable lower states~$\left|
1\right\rangle$ and~$\left| 3\right\rangle$. A field with the
frequency~$\omega_{a}$ couples the states $\left| 1\right\rangle$
and~$\left| 2\right\rangle$, while that with the
frequency~$\omega_{b}$ couples the states $\left| 3\right\rangle$
and~$\left| 2\right\rangle$ (see Fig.~1). The interaction of these
fields with atoms is described by the Rabi
frequencies~$\Omega_{a}={\boldsymbol{\mu}_{12}\cdot{\bf
E}_{a}}/{\hbar}$ and~$\Omega_{b}={\boldsymbol{\mu}_{32}\cdot{\bf
E}_{b}}/{\hbar}$, respectively. As the beams $a$ and $b$ are
spatially bounded, the Rabi frequencies depend on the position~${\bf
r}$ of an atom in space. The wave vectors will be considered close
in magnitude: $k_{a}\simeq k_{b}\simeq k$.

The equations for nondiagonal elements of the density matrix
in the rotating-wave approximation have the form
\[\dot{\rho}_{12}=i(\Delta_{a}-kv)\rho_{12}+\frac{i\Omega_{a}^{*}}{2}\,(\rho_{22}-\rho_{11})-\]
\begin{equation}
-\frac{i\Omega^{*}_{b}}{2}\rho_{13} +\left( \frac{\partial
\rho_{12}}{\partial t}\right)_{\mathrm{relax}} +\left(
\frac{\partial \rho_{12}}{\partial
t}\right)_{\mathrm{coll}}\,,\label{eq:rho12}
\end{equation}
\[\dot{\rho}_{23}=-i(\Delta_{b}-kv)\rho_{23}+\frac{i\Omega_{a}}{2}\rho_{13}+\]
\begin{equation}
+\frac{i\Omega_{b}}{2}\,(\rho_{33}-\rho_{22}) +\left(
\frac{\partial \rho_{23}}{\partial t}\right)_{\mathrm{relax}}
+\left( \frac{\partial \rho_{23}}{\partial
t}\right)_{\mathrm{coll}}\,,\label{eq:rho23}
\end{equation}
\[\dot{\rho}_{31}=-i(\Delta_{a}-\Delta_{b})\rho_{31}-\frac{i\Omega_{a}}{2}\rho^{*}_{23}
+\frac{i\Omega^{*}_{b}}{2}\rho^{*}_{12} +\]
\begin{equation}
+\left( \frac{\partial \rho_{31}}{\partial
t}\right)_{\mathrm{relax}} +\left( \frac{\partial
\rho_{31}}{\partial t}\right)_{\mathrm{coll}}.\label{eq:rho13}
\end{equation}
\noindent Here, the terms with the index ``$\mathrm{coll}$''
describe the relaxation processes due to collisions of active atoms
(interacting with the field) with atoms of the buffer gas
resulting in a change of the velocity, while the coherence is conserved.
The terms with the index ``$\mathrm{relax}$'' describe the rest
of relaxation processes, $\Delta_{a}=\omega_{21}-\omega_{a}$
and $\Delta_{b}=\omega_{23}-\omega_{b}$ are the one-photon
detunings, and $\Delta\omega=\Delta_{a}-\Delta_{b}$ is the
two-photon detuning. It is assumed that
\[ \left( \frac{\partial \rho_{12}}{\partial
t}\right)_{\mathrm{relax}}= -\Gamma_{12}\rho_{12}, \qquad\left(
\frac{\partial \rho_{23}}{\partial t}\right)_{\mathrm{relax}}=
-\Gamma_{23}\rho_{23},
\]
where $\Gamma_{12}$ and $\Gamma_{23}$ stand for the coherence
relaxation rates for the transitions $2\to 1$ and $2\to 3$,
respectively. If $\Gamma_{13}$ and $\Gamma_{23}$ are
sufficiently large, the collision terms in Eqs.~(\ref{eq:rho12})
and~(\ref{eq:rho23}) can be neglected. The relaxation rate for the
forbidden transition $|1\rangle\to|3\rangle$ is non-zero due to
collisions between atoms,
\begin{equation}
\left( \frac{\partial \rho_{31}}{\partial t}\right)_{\mathrm{relax}}
=-\gamma_{13}\rho_{31}.
\end{equation}

In the general case, the expression for the collision
term~\cite{sobelman2, sobelman3, rautian} can be presented as
\[
\left( \frac{\partial \rho_{ij}({\bf v}, {\bf v}\pr, t)}{\partial
t}\right)_{\mathrm{coll}} =-\nu\rho_{ij}({\bf v}, {\bf v}\pr, t)+\]
\begin{equation}
 +\int K_{ij}({\bf v}\pr, {\bf v})\rho_{ij}({\bf r},
{\bf v}\pr, t)\,d{\bf v}\pr\,, \label{collisional-term}
\end{equation}
where $K_{ij}({\bf v}\pr, {\bf v})$~ is the collision kernel, and $\nu$
is, in the general case, a complex-valued quantity with the frequency
dimension. It can be interpreted as the collision frequency in the
case where the scattering amplitudes in both states $i$ and $j$
are identical (see~\cite{sobelman2}).

The collision term can be simplified using the strong-collision
approximation (light atoms are scattered by heavy
particles~\cite{sobelman1}) and assuming that $\nu$ and $K_{ij}$ are
real, while the kernel $K_{ij}({\bf v}\pr, {\bf v})$ does not depend
on ${\bf v}\pr$, i.e. the velocity of an atom ${\bf v}$ after
a collision does not depend on its velocity~${\bf v}\pr$ before it. In
this case, the velocity distribution (arbitrary in the general case)
turns into the Maxwellian distribution after only several
collisions, that is, an atom quickly forgets its initial velocity.
As was shown in~\cite{sobelman1}, $K_{ij}({\bf v})=\nu W({\bf v})$,
where $W({\bf v})$ is the Maxwellian distribution.
Thus~\cite{rautian}, expression~(\ref{collisional-term}) takes the
form
\[
\left( \frac{\partial \rho_{ij}}{\partial t}\right)_{\mathrm{coll}}
=-\nu\bigl[ \rho_{ij}-W({\bf v})N_{ij}\bigr]\,,
\]
\begin{equation}
\mbox{where}\quad N_{ij}({\bf r}, t)=\int d{\bf v}\,\rho_{ij}({\bf
r}, {\bf v}, t)\, \label{n:definition}
\end{equation}
can be interpreted as the number of atoms with values of~$\rho_{ij}$
lying in the unit volume in the neighborhood of the point~${\bf r},$
and $\nu$ denotes the collision frequency~\cite{sobelman2}.

In this approximation, collisions with the buffer gas change only
the external degrees of freedom of atoms.

Using the approximation $\bigl|\frac{d}{dt}\bigr|\ll \Gamma_{12},
\Gamma_{23}$ and neglecting the collision terms
in~(\ref{eq:rho12}) and~(\ref{eq:rho23}), one can find the
stationary solutions for $\rho_{12}$ and $\rho_{23}$. After that,
with regard for the fact that the nondiagonal
elements~$\rho_{ii}$ are close to the equilibrium values
$\rho_{ii}^{(0)}$ (they differ from the latter by a small quantity
of the second order in the field intensity), one obtains
the equation for $\rho_{31}$:
\[
\dot{\rho}_{31} =-(\gamma_{13}+i\Delta\omega)\rho_{31} -\]
\[-\left[
\frac{\Omega_{a}^{*}\Omega_{a}}{\Gamma_{23}-i(\Delta_{b}-kv)}
+\frac{\Omega_{b}^{*}\Omega_{b}}{\Gamma_{12}+i(\Delta_{a}-kv)}
\right] \rho_{31}\]
\[ -\frac{\Omega_{a}\Omega_{b}^{*}}{4} \left[
\frac{\rho_{11}^{(0)}-\rho_{22}^{(0)}}{\Gamma_{12}+i(\Delta_{a}-kv)}
+\frac{\rho_{33}^{(0)}-\rho_{22}^{(0)}}{\Gamma_{23}-i(\Delta_{b}-kv)}
\right] +\]
\begin{equation}
+\left( \frac{\partial \rho_{31}}{\partial
t}\right)_{\mathrm{coll}}\,. \label{density-pre-main}
\end{equation}

\noindent The second term on the right-hand side describes the field
broadening. It can be neglected in the case of weak fields and
large relaxation rates~$\Gamma_{ij}$.

In order to simplify the equation, we also neglect the Doppler
broadening in the third term considering that $\Gamma_{ij}\gg
\Delta_{a, b}, \Delta\omega, kv$.

Taking into account that the equilibrium elements of the density
matrix~$\rho_{ii}$ are proportional to the distribution
function~$W({\bf v})$, we obtain a kinetic equation of the
Boltzmann type for~$\rho_{31}$.

In the stationary case of interest, it has the following form:
\[
({\bf v}\cdot\boldsymbol{\nabla})\rho({\bf r}, {\bf
v})=-(\nu+\gamma+i\Delta\omega)\rho({\bf r}, {\bf v})\]
\begin{equation}
+W({\bf v}) \bigl[ \lambda({\bf r})+ \nu N({\bf r}) \bigr].
\label{basic:equation}
\end{equation}

Here and below, we use the notations
\[
\rho({\bf r}, {\bf v})=\rho_{31}({\bf r}, {\bf v}),\qquad
\gamma=\gamma_{13},\]
\[N=N_{31},\qquad{}W({\bf v})=W_{0}e^{-{\bf v}\,^{2}/v_{0}^{2}},\]
where $N_{31}$ is determined by (\ref{n:definition}), and $W_{0}$ is the normalization constant of the Maxwellian distribution,
\begin{equation}
\lambda({\bf r})= \frac{1}{4}\left(
\frac{\rho_{11}^{(0)}}{\Gamma_{12}}+\frac{\rho_{33}^{(0)}}{\Gamma_{13}}
\right)\Omega_{a}\Omega_{b}^{*}\,. \label{eq:lambda}
\end{equation}
Here, we took into account that $\rho_{22}^{(0)}\ll\rho_{11}^{(0)}$,
$\rho_{22}^{(0)}\ll\rho_{33}^{(0)}$. As one can see
from~(\ref{eq:lambda}), the function $\lambda({\bf r})\sim
\Omega_{a}({\bf r})\Omega_{b}^{*}({\bf r})$ describes the transverse
profile of the beams.

For the sake of simplicity, we introduce the substitution
\[
\alpha_{0}=\gamma+i\Delta\omega\,,\quad
\alpha=\nu+\gamma+i\Delta\omega=\nu+\alpha_{0}\,.
\label{basic:signal}
\]
It is assumed that the collisions of atoms with walls result in the failure of the coherence between the lower states, so that
\[
\rho({\bf r}, {\bf v})|_{{\bf r}\in S}=0,
\]
where~$S$ denotes the surface confining the buffer-gas cell.

The shape of the spectral line is determined by the function
$T(\Delta\omega)=\re \left[S(\Delta\omega)/S(0)\right]$, where $S(\Delta\omega)$ has the form
\begin{equation}
S(\Delta\omega)=\iint d{\bf r}\,d{\bf v}\,\lambda({\bf r})\rho({\bf
r}, {\bf v}) =\int \lambda({\bf r})N({\bf r})\,d{\bf r}\,.
\label{basic:S-definition}
\end{equation}
In order to find it, it is necessary to obtain the function $\rho$.

In the case of a laser beam with the Gaussian intensity distribution
in the plane normal to the direction of its propagation, the
expression for $\lambda({\bf r})$ in the cylindrical coordinates
takes the form
\begin{equation}
\lambda(r, \varphi, z)=\lambda_{0}e^{-r^{2}/a^{2}}\,,\quad
\label{beams}
\end{equation}
where $\lambda_{0}$ is determined by~(\ref{eq:lambda}), and ${\bf r}$
lies at the beam axis.

In the strong-collision approximation, the time between
collisions~$\tau_{\nu}={1}/{\nu}$ is small as compared to the
characteristic time of flight of an atom through the interaction
region~$\tau_{a}={a}/{v_{0}}$. Therefore, we consider that
$\nu\tau_{a}\gg 1$.

In what follows, we investigate the effect of the beam size
and the distance to the walls of the gas cell on the lineshape
specified by the function $\re S(\Delta\omega)$. The form of this
function depending on the dimension of the problem will be
considered as well.


\section{Solution for Infinite Region}

\label{fourier-solution}

For an infinite region, Eq.~(\ref{basic:equation}) can be solved
with the help of the Fourier transformation with respect to the
argument ${\bf r}$:
\[
\hat{\rho}({\bf k}, {\bf v}) =\int\rho({\bf r}, {\bf v})\,e^{-i{\bf
k}\cdot{\bf r}}\,d{\bf r}\,,\quad \hat{N}({\bf
k})=\ds\int\hat{\rho}({\bf k}, {\bf v})\,d{\bf v}\,.
\]
Hence,
\begin{equation}
\hat{N}({\bf k})=\frac{\hat{\lambda}({\bf k})\hat{F}({\bf
k})}{1-\nu\hat{F}({\bf k})}\,,\quad \label{basic:N}
\end{equation}
where the function
\[
\hat{F}({\bf k})=\int\frac{W({\bf v})\,d{\bf v}}{\alpha+i{\bf k}{\bf
v}} =\frac{\sqrt{\pi}}{|k|v_{0}}\,
e^{\alpha^{2}/k^{2}v_{0}^{2}}\erfc \left(
\frac{\alpha}{|k|v_{0}}\right)\,,
\]
describes the Voigt profile. Let us find the signal $S_{\infty}$ using the properties of the Fourier transformation:
\begin{equation}
S_{\infty}(\Delta\omega) =\int d{\bf r}\,N({\bf r})\lambda({\bf r})
=\frac{1}{(2\pi)^{n}}\int d{\bf k}\,\hat{N}({\bf
k})\hat{\lambda}({\bf k})\,, \label{basic:infinite-S}
\end{equation}
where $n$ denotes the space dimension of the region (1, 2, or 3). Substituting $N$ from (\ref{basic:N}), we obtain
\[
S_{\infty}(\Delta\omega) =\frac{1}{(2\pi)^{n}}\int
\frac{\hat{\lambda}^{2}({\bf k})\hat{F}({\bf k})}{1-\nu\hat{F}({\bf
k})}\,d{\bf k}.
\]
In some cases, the value of $S(\Delta\omega)$ is mainly determined by small $k$ due
to the factor $\hat{\lambda}^{2}(k)$. That is why the function~$\hat{F}(k)$ can be replaced
by the asymptotic expansion in the quadratic approximation (see~\cite{abramovits}):
\begin{equation}
S_{\infty}(\Delta\omega) \simeq\frac{1}{(2\pi)^{n}}\int
\frac{\hat{\lambda}^{2}({\bf k})\,d{\bf
k}}{\alpha_{0}+\frac{k^{2}v_{0}^{2}}{2\alpha}}.
\label{basic:S-infty}
\end{equation}
It is valid for sufficiently small~$k$, for which
${|k|v_{0}}/{\nu}\ll 1$.

\subsection{One-dimensional case}
In the given case,~$\lambda(x)=\lambda_{0}e^{-x^{2}/a^{2}},$ and~(\ref{basic:S-infty}) has the form
\[
S^{(1)}_{\infty}(\Delta\omega) =\frac{\lambda_{0}^{2}a}{2}
\int\limits_{-\infty}^{\infty}
\frac{e^{-u^{2}/2}}{\alpha_{0}+u^{2}/\tau_{D}}\,du =\]
\begin{equation}
=\frac{\pi\lambda_{0}^{2}a^{3}}{2}\frac{\beta}{\alpha_{0}}\,e^{\beta^{2}a^{2}/2}\erfc\left(
\frac{\beta a}{\sqrt{2}}\right)\,, \label{diffusion:signal:1d}
\end{equation}
where
\[
\beta^{2}=\frac{2\alpha\alpha_{0}}{v_{0}^{2}}\,,\quad
\tau_{D}=\frac{2\alpha a^{2}}{v_{0}^{2}}\,,\quad
\beta^{2}a^{2}=\tau_{D}\alpha_{0}
\]
(the quantity $\tau_{D}$ will be interpreted below).
In our approximation, $\nu\gg\gamma, \Delta\omega$, that is why one can consider $\tau_{D}=\const$.
The right-hand side of~(\ref{diffusion:signal:1d}) can be presented in the form of a superposition of the profiles
\begin{equation}
S^{(1)}_{\infty}(\Delta\omega)
\sim\int\limits_{-\infty}^{\infty}s(\Delta\omega, u)g^{(1)}_{\infty}(u)\,du\,,\quad
\label{lorentzian}
\end{equation}
where
\[
s(\Delta\omega, u)
=\frac{\gamma_{\eff}(u)}{\gamma_{\eff}(u)+i\Delta\omega}\,,\quad
\gamma_{\eff}(u)=\gamma+u^{2}/\tau_{D}
\]
with the weighting factor (independent of $\omega$)
\begin{equation}
g^{(1)}_{\infty}(u)=\frac{1}{\gamma_{\eff}(u)}\,e^{-u^{2}/2}\,.
\label{weight:inf:1g}
\end{equation}
The combination $\tau(u)=\tau_{D}/u^{2}$ can be called the
effective diffusion time of an atom; $\tau(u)$ for ``fast'' atoms
is larger than that for ``slow'' ones.

The profile $S^{(1)}_{\infty}(\Delta\omega)$ can be described as
the effective Lorentzian with the center at the origin of
coordinates
\[
S_{L}(\Delta\omega)=\frac{\Gamma_{0}}{\Gamma_{0}+i\Delta\omega}.
\]
Its width $\Gamma_{0}$ is determined by the relation
\[
\Gamma_{0}^{2}=-\frac{2S(0)}{S''(0)}.
\]
Simple calculations for $\gamma\tau_{D}\ll 1$ yield
\begin{equation}
\Gamma_{0}^{2}=\dfrac{8}{3}\gamma^{2}\cdot
\left( 1-\sqrt{\dfrac{2}{\pi}}\,\sqrt{\gamma\tau_{D}}
+\dots\right)\,.
\label{diffusion1:gamma1}
\end{equation}
Therefore, we obtain $\Gamma_{0}\simeq \sqrt{\frac{8}{3}}\,\gamma$ for small $\gamma$.

\subsection{Two-dimensional case}
Using the similar procedure for the function~$\lambda({\bf
r})=\lambda_{0}e^{-r^{2}/a^{2}}$ in the polar coordinates, we
obtain
\[
S^{(2)}_{\infty}(\Delta\omega) =\frac{\pi\lambda_{0}^{2}a^{2}}{2}
\int\limits_{0}^{\infty}
\frac{ue^{-u^{2}/2}\,du}{\alpha_{0}+u^{2}/\tau_{D}}=\]
\begin{equation}
 =\frac{\pi
\lambda_{0}^{2}a^{4}}{4\alpha_{0}}\,e^{\beta^{2}a^{2}/2}\ei_{1}\left(
\frac{\beta^{2}a^{2}}{2}\right)\,. \label{diffusion:signal:2d}
\end{equation}
Here, $\ei_{1}$ denotes the integral first-order
exponent\footnote{Notation:
$\ds\ei_{1}(x)=\int\limits_{x}^{\infty}\frac{e^{-t}}{t}\,dt$\,.}.
The obtained expression is similar to~(\ref{diffusion:signal:1d})
and has the same interpretation, though with the weighting function
\begin{equation}
g^{(2)}_{\infty}(u)=\frac{1}{\gamma_{\eff}(u)}\,ue^{-u^{2}/2}\,.\quad
\label{weight:inf:2g}
\end{equation}
The width of the effective Lorentzian for $\gamma\tau_{D}\ll 1$ has the form
\begin{equation}
\Gamma_{0}^{2}=\gamma^{2} 2\bigl[\ln 2-\gamma_{E}-\ln
(\gamma\tau_{D})\bigr]+o(\gamma\tau_{D})\,.
\label{diffusion1:gamma2}
\end{equation}


\section{Effective Diffusion Equation}
\label{diffusion1}

\subsection{Derivation of diffusion equation}
According to~(\ref{basic:S-definition}), the complex
signal~$S(\Delta\omega)$ can be expressed with the help of the
zero-order momentum $N({\bf r})$ of the distribution $\rho({\bf r})$
with respect to ${\bf v}$. Let us find the equation for the moments
of higher orders $N^{(k)}({\bf r})$.

First, we consider the one-dimensional case. Here,
\[
N^{(k)}(x)=\int\limits_{-\infty}^{\infty}v^{k}\rho(x,
v)\,dv\,,\quad N^{(0)}(x)=N(x)\,.
\]
Equation~(\ref{basic:equation}) presented in the one-dimensional case as
\begin{equation}
v\frac{\partial \rho(x, v)}{\partial x}=-\alpha\rho(x,
v)+W(v)\bigl[ \lambda(x)+\nu N^{(0)}(x)\bigr]
\label{equation:diff:1d}
\end{equation}
will be multiplied by $v^{k}$ and integrated over $v$. This
procedure yields a chain of equations for the moments
$N^{(k)}(x)$.

Writing down the obtained equations separately for even and odd
$k$ and successively substituting the following equation into the
previous one~$m$ times, we obtain the following expression for
$N^{(0)}$:
\[
\alpha N^{(0)}= \sum\limits_{k=0}^{m} \frac{\langle
v^{2k}\rangle}{\alpha^{2k}}\frac{d^{2k}}{dx^{2k}}\bigl[
\lambda(x)+\nu N^{(0)}(x)\bigr]+\]
\begin{equation}
+\frac{1}{\alpha^{2m+1}}\frac{d^{2m+2}N^{(2m+2)}}{dx^{2m+2}} \,.
\label{basic:expansion1}
\end{equation}
The result for higher space dimensions will be similar, though
more complicated as the moments will be specified by tensors. A
similar procedure yields
\[
\alpha N^{(0)}({\bf r}) =\sum\limits_{k=0}^{m} \frac{\langle
v_{i_{1}}\dots v_{i_{2k}}\rangle}{\alpha^{2k}}\times\]
\begin{equation}
\times\bigl(\nabla_{i_{1}}\dots\nabla_{i_{2k}}\bigr)
\bigl[\lambda({\bf r})+\nu N^{(0)}({\bf r})\bigr]+\dots
\label{basic:expansion2}
\end{equation}
Expressions~(\ref{basic:expansion1}) and~(\ref{basic:expansion2})
represent asymptotic expansions, where one can leave only the
first terms in the case of sufficiently large $\nu$ (and
$\alpha$). In particular, the second-order terms result in the
diffusion equation.

Using the known expressions for the averages $\langle
v_{i}v_{j}\rangle$, one obtains
 \[
N^{(0)}= \left( 1+\frac{\langle
v^{2}\rangle}{n\alpha^{2}}\,\laplace \right)(\lambda+\nu
N^{(0)})+\dots\,
\]
in the quadratic approximation, where $n$ stands for the space dimension. In view of the
equality $\langle v^{2}\rangle=\frac{n}{2}\,v_{0}^{2}$, one
derives
\begin{equation}
\alpha_{0} N({\bf r})= \frac{\nu v_{0}^{2}}{2\alpha^{2}}\,\Delta
N({\bf r})+\lambda({\bf r})\,,\quad N(\infty)=0\,
\label{diffusion:equation}
\end{equation}
for the arbitrary~$n$ (accurate to terms of the order
of~$1/\nu$). This equation can be interpreted as the diffusion equation with
the absorption coefficient~$\alpha_{0}=\gamma+i\Delta\omega$ and
the complex-valued diffusion coefficient
\begin{equation}
\tilde{D}=\frac{\nu v_{0}^{2}}{2\alpha^{2}}.
\label{diffusion:coeffc}
\end{equation}
The diffusion coefficient is identical for all space dimensions.

Expression~(\ref{diffusion:coeffc}) at large $\nu$ becomes real and turns into
\begin{equation}
D=\frac{v_{0}^{2}}{2\nu}.
\label{diffusion:coeff}
\end{equation}
Proceeding from the formula $a=\sqrt{D\tau_{D}}$, we obtain the
characteristic diffusion time (approximate time, for which an atom
leaves the beam)
\begin{equation}
\tau_{D}=\frac{a^{2}}{D}=\frac{2a^{2}\nu}{v_{0}^{2}}.
\label{diffusion:time}
\end{equation}

For the further consideration, it is convenient to put down the
diffusion equation~(\ref{diffusion:equation}) in the form
\begin{equation}
\Delta N({\bf r})-\beta^{2}N({\bf r})=-f({\bf r})\,,
\label{diffusion:equation2}
\end{equation}
where $ f({\bf r})=\frac{\beta^{2}}{\alpha_{0}}\,\lambda({\bf
r}),$
 $\beta^{2}=\frac{2\alpha_{0}\alpha^{2}}{\nu v_{0}^{2}}
\simeq \frac{\alpha_{0}\tau_{D}}{a^{2}}.$  Solving it with the help
of the Green function method, one can see that the expression for
the signal obtained by solving~(\ref{diffusion:equation2}) in the case
of large~$\nu$ will be the same as that obtained earlier by direct
calculations~(\ref{basic:S-infty}). One can also see that the number
of terms used in the asymptotic expansion $\hat{F}({\bf k})$ of
expression~(\ref{basic:S-infty}) correlates with the number of terms
of Eq.~(\ref{basic:expansion2}) that must be kept in order to obtain
the diffusion equation.

In the case of a finite cell, the kinetic
equation~(\ref{basic:equation}) can be solved formally, by
interpreting the last term on the right-hand side as a
nonuniform one (see~\cite{mors}). The result will be the same as
that derived from the solution of~(\ref{diffusion:equation})
accurate to the terms $\nu^{-2}$ (to which the diffusion equation
is actually valid).

The atomic motion is characterized by five characteristic times:
\[
\tau_{a}=\frac{a}{v_{0}}\,,\quad \tau_{R}=\frac{R}{v_{0}}\,,\quad
\tau_{\gamma}=\frac{1}{\gamma}\,,\]
\begin{equation}
\tau_{\nu}=\frac{1}{\nu}\,,\quad
\tau_{D}=\frac{a^{2}}{D}=\nu\tau_{a}^{2}\,.
\label{diffusion:times}
\end{equation}
According to the accepted approximations, they satisfy the following conditions:
\begin{equation}
\tau_{\nu}<\tau_{a}<\tau_{R}\,,\quad
\tau_{D}<\tau_{\gamma}\,,\quad
\tau_{\nu}\ll\tau_{\gamma}\,,\quad
\tau_{a}<\tau_{D}\,.
\label{diffusion:times:conditions}
\end{equation}
The first condition results from the strong-collision
approximation and the geometric configuration~$R>a$ for
small~$\nu$ and (or) large~$R$. The approximation used for the
derivation of the diffusion equation will be valid for a
finite-size cell. The second and third conditions correspond
to the slowness of relaxation processes, while the last condition
is evident.

Let us introduce the dimensionless time and space scales
$\hat{t}=\gamma t$ and $\hat{r}=r/a$ and denote the dimensionless
velocity by~$\hat{v}_{0}={v_{0}}/{(\gamma a)}$, the diffusion
coefficient by~$\hat{D}={D}/{(\gamma a^{2})}$, the collision
frequency by~$\hat{\nu}={\nu}/{\gamma}$, and the cell size
by~${\hat{R}=R/a}$. The dimensionless characteristic times will
have the form
\[ \hat{\tau}_{\gamma}=1\,,\quad
\hat{\tau}_{\nu}=\gamma\tau_{\nu}=\frac{1}{\hat{\nu}}\,,\quad
\hat{\tau}_{D}=\gamma\tau_{D}=\frac{2\hat{\nu}}{\hat{v}_{0}^{2}}\,,\]
\begin{equation}
\hat{\tau}_{a}=\gamma\tau_{a}=\frac{1}{\hat{v}_{0}}\,,\quad
\hat{\tau}_{R}=\gamma\tau_{R}=\frac{\hat{R}}{\hat{v}_{0}}\,.
\label{diffusion:times:dimless}
\end{equation}
As will be seen from the further solution, the shape of the
transmission line in the infinite case is completely determined by
two dimensionless characteristic times~$\hat{\tau}_{\nu}$
and~$\hat{\tau}_{D}$, whereas, in the case of a finite cell, it depends on
its size~-- $\hat{R}$ (width) and $\hat{l}$ (length, in the
three-dimensional case).

The diffusion equation in the form~(\ref{diffusion:equation2}) will be solved for
different space dimensions. The most effective approach is that based on the Green functions.
In order to simplify the comparison of the cases of finite $R$ and $R\to\infty$,
we present the result of investigating the effect of space confinements on
the signal in the form most similar to the solution for an infinite cell.

\subsection{One-dimensional case}
Using the general solution of the one-dimensional diffusion equation with the
boundary conditions $N(\pm R)=0$, we obtain
\begin{equation}
S^{(1)}_{R}(\Delta\omega)=\frac{1}{2\pi a}\int\limits_{-\infty}^{+\infty}
\frac{\hat{\lambda}(u/a)\hat{\lambda}_{R}(u/a)\,du}{\alpha_{0}+u^{2}/\tau_{D}}\,,
\label{signal-1dR}
\end{equation}
where $\hat{\lambda}_{R}(u/a)=\hat{\lambda}(u/R)b^{(1)}(u, R),$ and
$b^{(1)}(u, R)$ is a factor depending on $R$:
\[b^{(1)}(u, R)=B^{(1)}(u/a, R)\,,\]
\begin{equation}
 B^{(1)}(k,
R)=\frac{1}{\hat{\lambda}(k)} \int\limits_{-R}^{R}\lambda(x)
\left[ e^{-ikx}-e^{-ikR}\frac{\ch(\beta x)}{\ch(\beta R)}
\right]\,dx . \label{signal-1B}
\end{equation}
This expression means that, similar to the infinite case~(\ref{lorentzian}),
$S^{(1)}_{R}$ can be written down as a superposition of Lorentzians:
\[
S^{(1)}_{R}(\Delta\omega)\sim \int\limits_{-\infty}^{+\infty}
s(\Delta\omega, u)g^{(1)}_{R}(u, \Delta\omega)\,du\,,
\]
though, in contrast to~(\ref{weight:inf:1g}), the weighting
function in the case of finite $R$ depends on~$R$
and~$\Delta\omega$:
\[
g^{(1)}_{R}(u, \Delta\omega)=g^{(1)}_{\infty}(u)\,b^{(1)}(u,
\Delta\omega, R)\,;
\]
moreover, $\lim_{R\to\infty}g^{(1)}_{R}(u,
\Delta\omega)=g^{(1)}_{\infty}(u)$. In the case of $R\to\infty$, the
dependence on $\Delta\omega$ disappears. The factor $g^{(1)}_{R}(u,
\Delta\omega)$ cannot be interpreted as a weighing one due to the
dependence on $\Delta\omega$. In addition, the Lorentzian
$s(\Delta\omega, u)$ cannot be replaced by the more complicated
profile $s_{R}(\Delta\omega, u)=s(\Delta\omega, u)\cdot b^{(1)}(u,
\Delta\omega, R)$ in order to separate out the weight, as it was
done in the infinite case, because this ``profile'' $s_{R}$ becomes
singular for some values of $u$ (in particular, for $u\to\infty$),
though this singularity is compensated by the other factor
$g^{(1)}_{\infty}(u)$. Thus, the interpretation of
$S^{(1)}_{R}(\Delta\omega)$ as a weighted superposition is
impossible here.

For the Gaussian intensity distribution, the function $b^{(1)}(k, R)$ can be expressed
in terms of the error functions:
\[
B^{(1, g)}(k, R)=1-
\frac{\erfc\left(\frac{R}{a}+i\frac{ka}{2}\right)+\erfc\left(\frac{R}{a}-i\frac{ka}{2}\right)}{2}-\]
\begin{equation}
-e^{(k^{2}+\beta^{2})/4} \frac{\cos kR}{\ch \beta R}
\frac{\erf\left(\frac{R}{a}+\frac{\beta
a}{2}\right)+\erfc\left(\frac{R}{a}-\frac{\beta
a}{2}\right)}{2}\,. \label{signal-1B-gauss}
\end{equation}
In the case $R\gg a$, $\frac{R}{a}\gg \frac{\beta a}{2}$, one can obtain the asymptotic
behavior of $B^{(1)}(k, R)$ for small~$k$:
\begin{equation}
B^{(1, g)}(k, R)\simeq
1-e^{(k^{2}+\beta^{2})a^{2}/4}\,\frac{\cos kR}{\ch \beta R}\,.
\label{diffusion2:b:asympt:1d}
\end{equation}

The graphs of $\re S_{R}(\Delta\omega)$ and the associated
parameters for the typical values
\[
\gamma=1.0\times 10^{2}\; \text{Hz}\,,\quad \nu=1.0\times 10^{6}\;
\text{Hz}\,,\]
\[ a=1.0\times 10^{-3}\; \text{m}\,,\quad
v_{0}=2.95\times 10^{2}\;\text{m/s} ,
\]
and $\tau_{D}=2.3\times10^{-5}\;\text{s}$ with the corresponding
dimensionless values
\[
\hat{\tau}_{D}=2.3\times10^{-3}\,,\quad
\hat{\tau}_{\nu}=1.0\times10^{-4}
\]
are given in Fig.~2.

One can see that the line will be narrower for larger~$R$ (where
the probability to return without loss of coherence is higher).
Formally, the geometric factor ~$g^{(1)}_{R}(u, \Delta\omega)$
for finite $R$ is larger than~$g^{(1)}_{\infty}(u)$. Therefore, the
profile $\re S_{R}(\Delta\omega)$ will be wider.

\subsection{Two-dimensional case}

Using the axially symmetric solution of the two-dimensional
diffusion equation in the region $r<R$ with the boundary condition
$N({\bf r})\bigr|_{\,r=R}=0$, one obtains
\begin{equation}
S^{(2)}_{R}(\Delta\omega)=
\frac{1}{2\pi}\int\limits_{0}^{\infty}\frac{u\hat{\lambda}(u/a)\hat{\lambda}_{R}(u/a)\,du}{\alpha_{0}+u^{2}/\tau_{D}}\,b^{(2)}(u, R)\,,
\label{profile-2dR-S}
\end{equation}
where $\hat{\lambda}_{R}(u/a)=\hat{\lambda}(u/a)b^{(2)}(u, R)$ and
\[b^{(2)}(u, R)=B^{(2)}(u/a, R)\,,\]
\begin{equation}
B^{(2)}(k, R)\!=\!
\frac{2\pi}{\hat{\lambda}(k)}\!\int\limits_{0}^{R}r\lambda(r)\!\left[
J_{0}(kr)\!-\!J_{0}(kR)\frac{I_{0}(\beta r)}{I_{0}(\beta R)}
\right]\!dr\,; \label{diffusion2:signal:2d}
\end{equation}
moreover, $\lim_{R\to\infty}b^{(2)}_{R}(u, R)=1$.

The same way as in the one-dimensional case, the weighting factor of the Lorentzian has the form
$g^{(2)}_{R}(u, \Delta\omega)=g^{(2)}_{\infty}(u)b^{(2)}(u, R, \Delta\omega)$.
The interpretation of~(\ref{diffusion2:signal:2d}) and the comparison with the infinite
case~(\ref{diffusion:signal:2d}) are the same as those in the one-dimensional one~(\ref{signal-1B-gauss}).
The graphs of $\re S^{(2)}_{R}(\Delta\omega)$ in Fig.~3 are plotted for the same values of
the parameters as in Fig.~2.

\begin{figure}
\includegraphics[width=87mm]{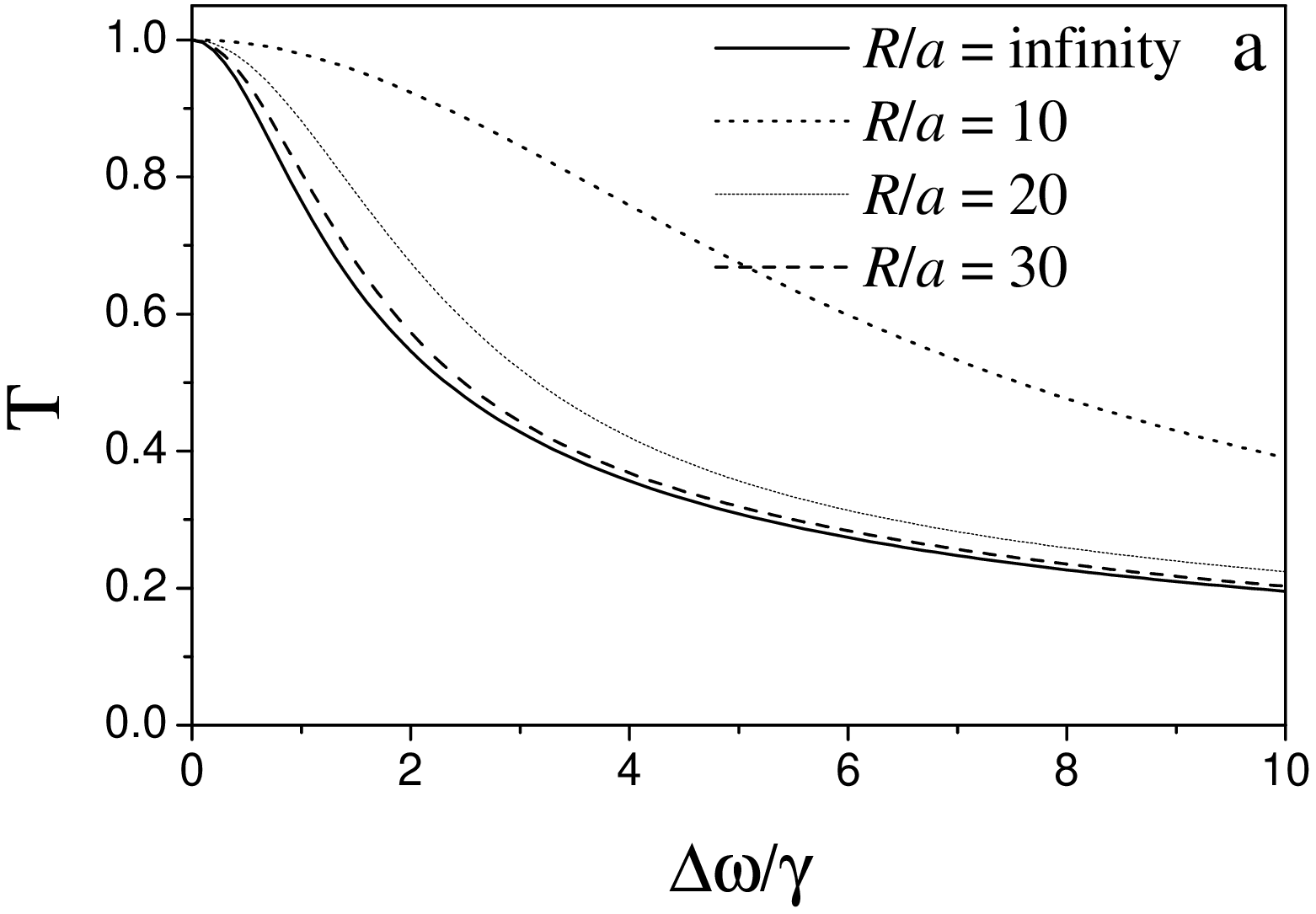}\vskip2mm
\includegraphics[width=87mm]{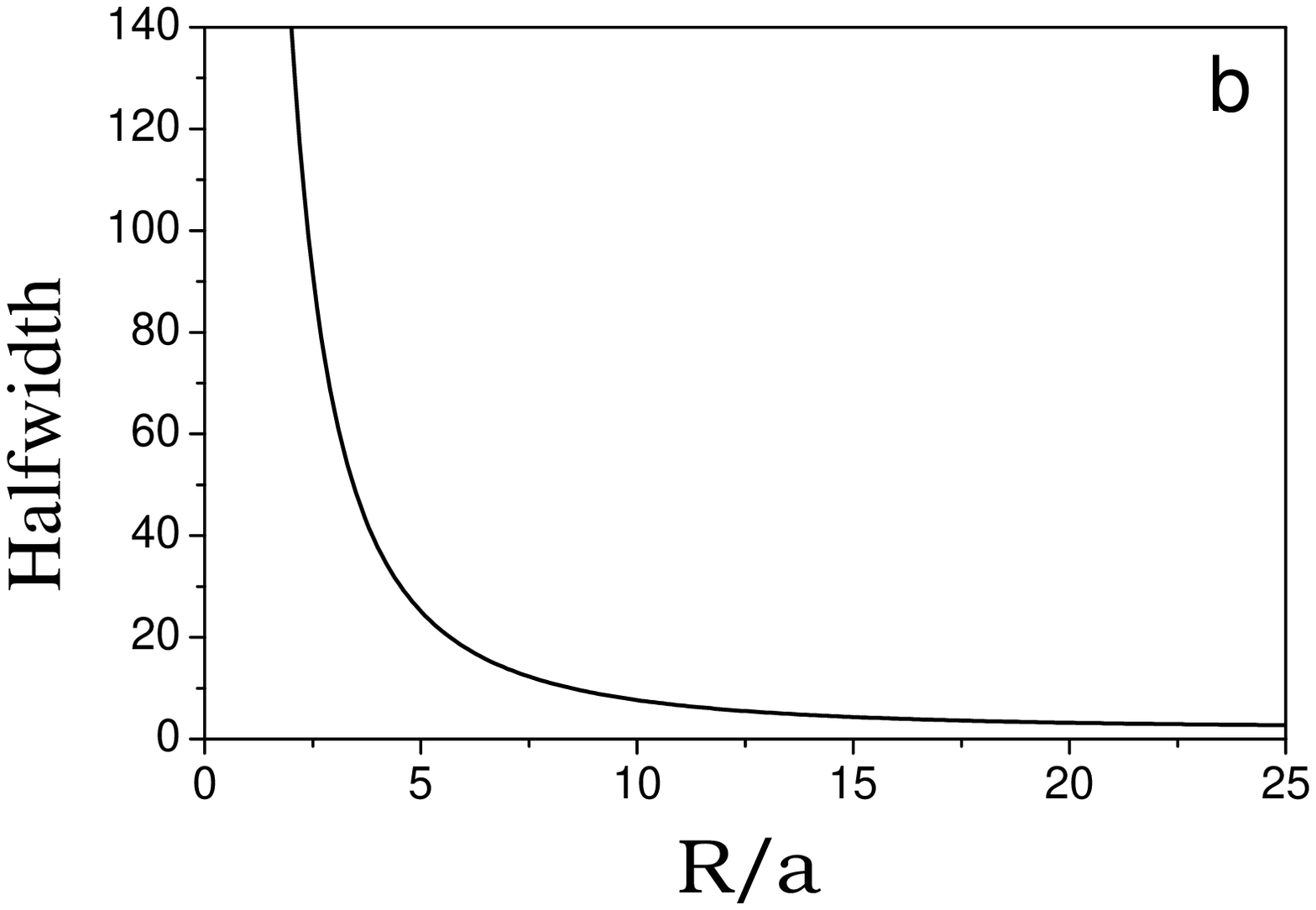}
\caption{One-dimensional case: normalized transmission $T=\re
S(\Delta\omega)/S(0)$ as a function of $\Delta\omega/\gamma$ for
different~$R$ ({\it a}) and the transmission spectrum halfwidth as a
function of~$R/a$ ({\it b}). The values of the parameters are given in
text  }
\end{figure}

\begin{figure}
\includegraphics[width=87mm]{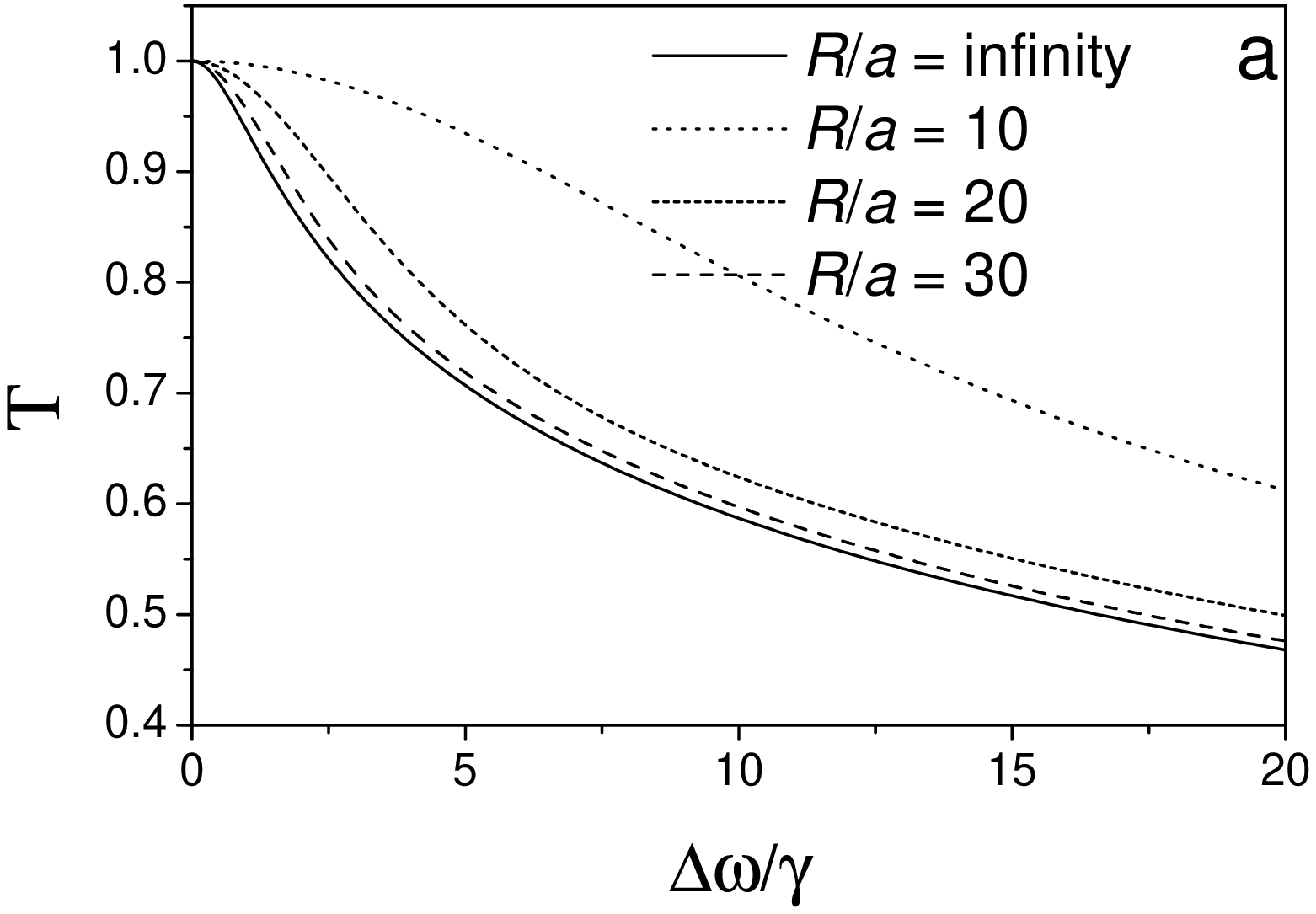}\vskip2mm
\includegraphics[width=87mm]{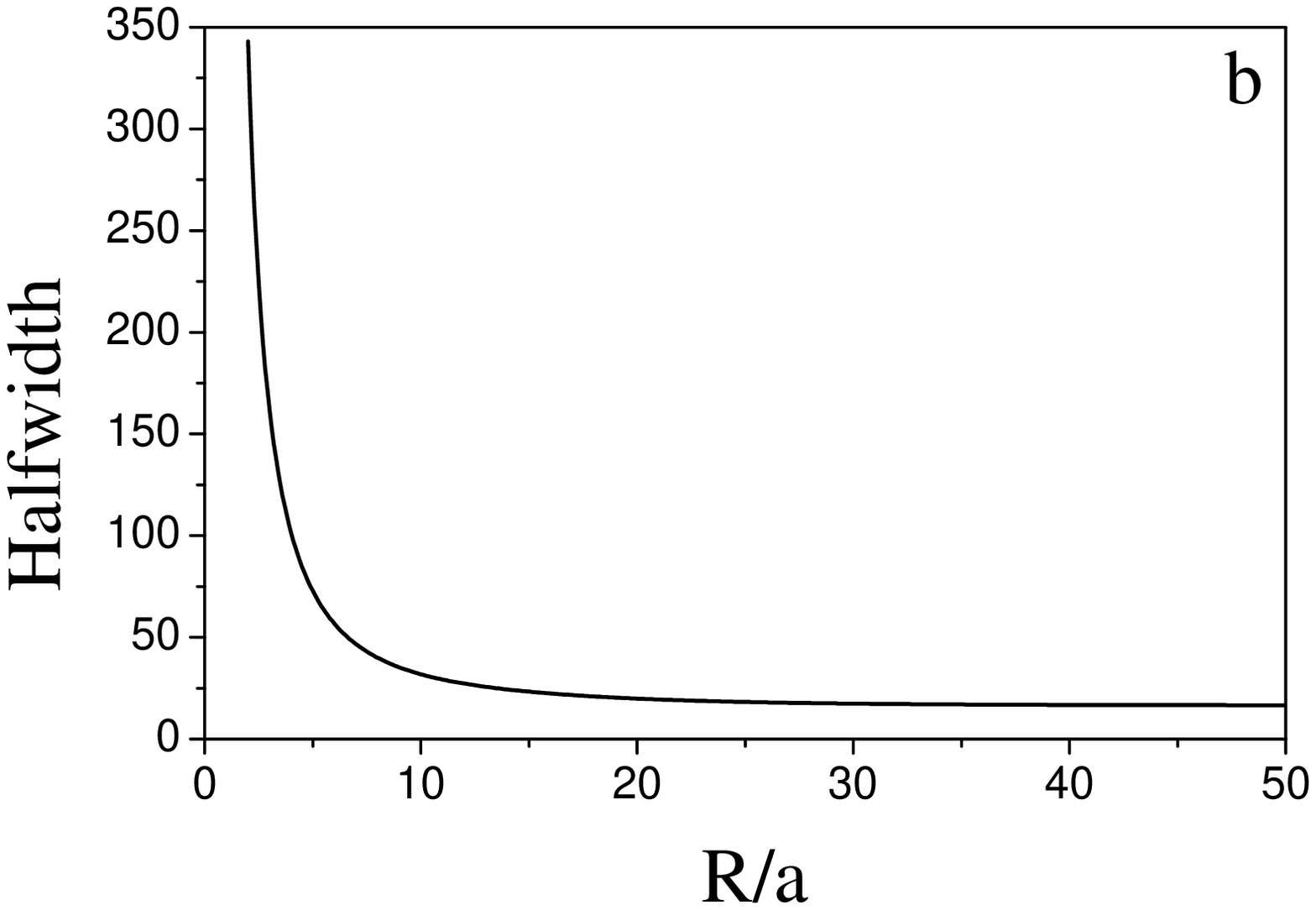}
\caption{Two-dimensional case: normalized transmission $T=\re
S_{R}(\Delta\omega)/S_{R}(0)$ as a function of
$\Delta\omega/\gamma$ for~$R=\infty$ and finite $R$ ({\it a}) and
the transmission spectrum halfwidth as a function of~$R/a$ ({\it b}). The
values of the parameters are the same as in Fig. 2  }
\end{figure}

\subsection{Three-dimensional case}
\label{signal3d}

In the three-dimensional case where~$N({\bf r})$ is zero on the
cylindrical surface of radius~$r=R$ confined by the planes~$z=\pm
l$, the diffusion equation is identical to that obtained in the
two-dimensional case, whereas the signal has the following form
(modified two-dimensional profile):
\[
S^{(3)}_{R}(\Delta\omega) =\frac{l}{16\pi
a^{2}}\frac{\beta^{2}}{\alpha_{0}}\times\]
\begin{equation}
\times\int\limits_{0}^{\infty}du\,
\frac{u\hat{\lambda}^{2}(u/a)}{\alpha_{0}+u^{2}/\tau_{D}}\,b^{(2)}(u,
R)b^{(3)}(u, l)\,, \label{signal3d-fin-S}
\end{equation}
where $\hat{\lambda}(k)$ denotes the two-dimensional Fourier
transform of $\lambda({\bf r})$, $b^{(2)}(u, R)$ is the factor
from the two-dimensional case depending on~$R$
(see~(\ref{diffusion2:signal:2d})), and
\[b^{(3)}(u, l, \Delta\omega)=1-\frac{\tanh\xi(u,
\Delta\omega)}{\xi(u, \Delta\omega)}\,,\]
\begin{equation}
\xi(u, \Delta\omega)=\frac{l}{a}\sqrt{u^2+\beta^{2}a^{2}}\,.
\label{signal3d-fin-B-l}
\end{equation}
The weighting factor of the Lorentzians in the three-dimensional case~(\ref{signal3d-fin-S}) has the form
\[
g^{(3)}_{R, l}(u, \Delta\omega)=g^{(2)}_{R}(u) b^{(3)}(u, l)=\]
\[ =g^{(2)}_{\infty}(u) b^{(2)}(u, R, \Delta\omega)
b^{(3)}(u, l, \Delta\omega).
\]
Its properties are similar to those in the one- and two-dimensional cases. It is worth noting
that its dependence on the cell dimensions~$R$ and~$l$ appears in the form of independent factors.

\begin{figure}
\includegraphics[width=7.8cm]{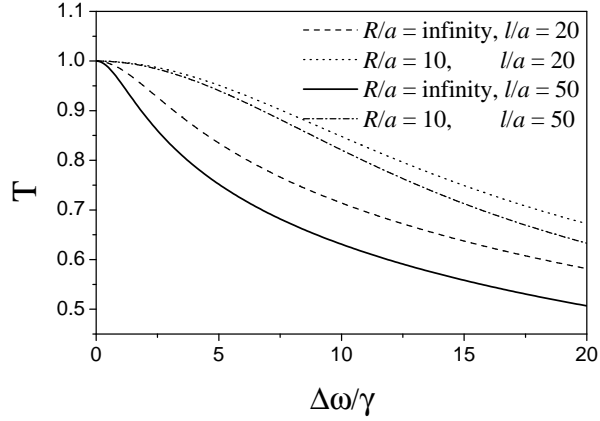}
\caption{Three-dimensional case. $T=\re
S^{(3)}_{R}(\Delta\omega)/S^{(3)}_{R}(0)$ as a function of
 $\Delta\omega/\gamma$ for different space parameters  }\vskip3mm
\end{figure}

\begin{figure}
\includegraphics[width=8.0cm]{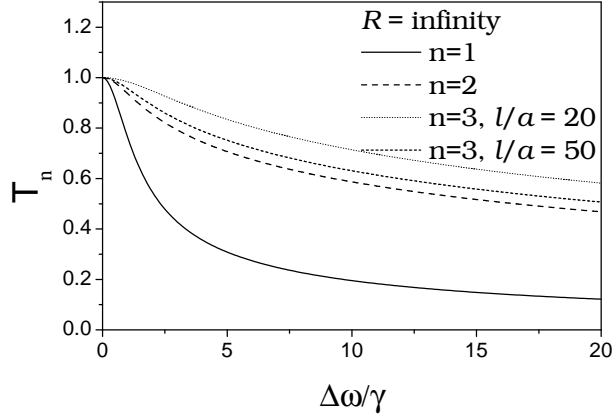}
\caption{Comparison of the functions $T_{n}=\re
S^{(n)}_{\infty}(\Delta\omega)/S^{(n)}_{\infty}(0)$ for $n=1,2,3$ }
\end{figure}

The transmission spectra normalized to one at the maximum for
different values of the space parameters are given in Fig.~4. As
expected, a decrease of $l$ results in the broadening of the
spectrum. Figures~5 and~6 present the transmission spectra for all
dimensions in the case of an infinite cell and a cell with the
size $R/a=10$. One can see that the narrowest spectrum is obtained
in the one-dimensional case and the widest -- in the
three-dimensional one. The values of the parameters used for
plotting the graphs in Figs.~4--6 are the same as in Fig.~2.

\begin{figure}
\includegraphics[width=87mm]{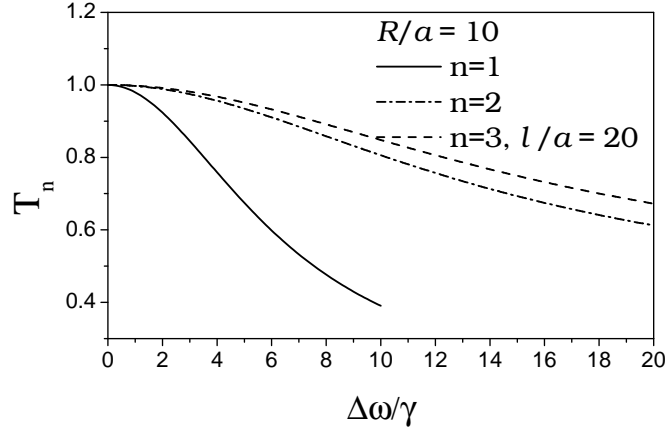}
\caption{Functions $T_{n}=\re S^{(n)}_{R}(\Delta\omega)/
S^{(n)}_{R}(0)$, $n=1,2,3$, for $R/a=10$  }\vskip3mm
\end{figure}

\begin{figure}
\includegraphics[width=8cm]{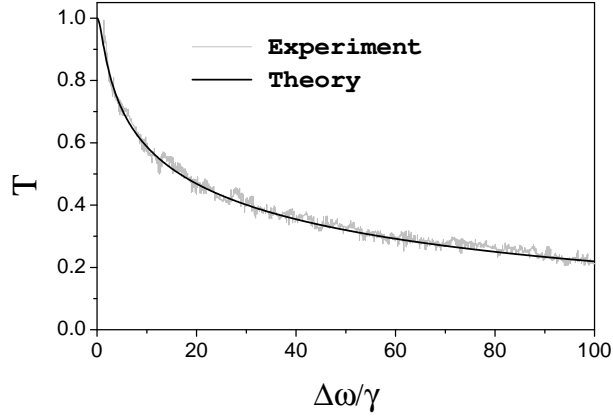}
\caption{Normalized transmission $T=\re
S^{(3)}_{R}(\Delta\omega)/S^{(3)}_{R}(0)$~--- comparison with
experiment  }
\end{figure}

\section{Comparison with Experiment}

Figure 7 shows the comparison of the results of numerical calculations with
experimental data~\cite{xiao}. The calculations were performed for
the dimensionless parameters
\begin{gather*}
\hat{\tau}_{a} = 3.39\times 10^{-4}\,,~~~~ \hat{\tau}_{r} =
1.69\times 10^{-3}\,,
\\\tau_{\nu}     = 1.00\times 10^{-4}\,,~~~~
\hat{\tau}_{D} = 2.30\times 10^{-3}\,,\\ \hat{v}_0  = 2.95\times
10^{3}\,,~~~~~~
\hat{D}    = 4.35\times 10^{2}\,,\\
\hat{R}    = 5\,,~~~~ \hat{l}    = 5\,
\end{gather*}
close to the experimental conditions. One can see that our
theoretical results are in rather good agreement with those of
experimental measurements.

It is worth noting that the experimental data presented in Fig.~7
also agree well with the theoretical calculation of the
transmission spectrum obtained by averaging over atom
trajectories. This testifies to the fact that the proposed model is
also in good agreement with that based on the averaging over
trajectories~\cite{xiao,xiao2008}, at least in the cases where
these theories can be consistently compared. The proposed model
has a wider range of applications, since the model with averaging over
trajectories can be used only for the case of an infinite cell (at
least in its present form).


\section{Conclusions}
\label{summary}

We have constructed a model for the description of the phenomenon of
Ramsey diffusion-induced narrowing of the transmission spectrum of
atoms in a buffer-gas cell for the case of weak fields in the
strong-collision approximation. This model can be used for an
arbitrary intensity distribution of laser beams in the direction
normal to their propagation. The general theory is illustrated by
calculations for the Gaussian intensity distribution.

The analytical expressions for transmission spectra obtained with
the help of the effective diffusion equation qualitatively agree
with experimental data and the results obtained by averaging over
atom trajectories. The proposed model gives a possibility to
investigate the shape of the transmission spectrum as a function of
not only the intensity distribution in the plane of the laser beam
but also the size of the buffer-gas cell. We have considered different
geometric configurations (one-, two-, and three-dimensional).
Comparing the spectra for \mbox{one-,} two-, and three-dimensional
models, one can see that the line becomes wider with increase in the
dimension.

\vskip3mm The work is carried out in the framework of Projects
F28.2/035 and RFFD/1-09-25.


\end{document}